# Anisotropic AC Behavior of Multifilamentary Bi-2223/Ag Tapes

J.-F. Fagnard, P. Vanderbemden, R. Cloots, and M. Ausloos

*Abstract*—In this communication, we report on the anisotropy of the superconducting properties of multifilamentary Bi-based tapes experimentally investigated by AC magnetic susceptibility measurements. The susceptibility $\chi = \chi' - j\chi''$ was measured using a commercial system and a couple of orthogonal pick-up coils. The $\chi''$ vs. temperature curves were shown to exhibit two peaks. The smaller of the peaks, occurring near $T = 72$ K, was only visible for particular field directions and within a given frequency window. Such results point out the role played by the phase difference between the applied magnetic field and the internal magnetic field seen by the filaments.

*Index Terms*—AC susceptibility, Bi-2223/Ag multifilamentary tape, magnetic field orientation.

## I. INTRODUCTION

THE characterization of magnetic behavior of HTS tapes can be carried out by several techniques, including transport and magnetic measurements. The present work focuses on AC susceptibility measurements.

Generally speaking, the AC magnetic response of a Bi-2223/Ag tape is given by a combination of superconducting shielding currents in the filaments and eddy currents in the silver matrix [1]–[5]. Recently, an extensive study of the AC susceptibility of monocore tapes has been carried out by Savvides and Müller [4]. From these investigations, the following considerations should be kept in mind: (i) the eddy currents are predominant when the skin depth is comparable or smaller than the thickness of silver, i.e., at high frequencies and low temperatures, (ii) unlike the superconducting behavior, the eddy current susceptibility is not directly affected by the AC field amplitude, (iii) for the superconducting core, the frequency effects (modifying the flux creep) are usually much smaller than the field dependence of the critical current.

In the case of multifilamentary tapes, additional 'hybrid' shielding loops (interfilament coupling) consisting of superconducting and resistive current paths are present. This leads to a modification of the AC losses mechanisms [6], [7]. The AC magnetic response is thus quite intricate because of the complex geometries involved. A reduction of the AC losses is generally achieved by twisting the filaments; in HTS tapes however, filaments are often untwisted.

In this paper, we have investigated the AC magnetic susceptibility of Bi-2223/Ag multifilamentary tapes with the emphasis placed on studying the influence of the magnetic field orientation when the field is parallel to the main tape plane.

## II. EXPERIMENT

Several samples of dimensions 0.18 mm × 3.35 mm × 3.4 mm were cut from a Bi-2223/Ag multifilamentary tape from Nordic Superconductor Technologies. The tape contains 38 untwisted filaments whose average cross-section is estimated to be 300 $\mu$m × 22 $\mu$m. The grains are textured in such a way that their $c$-axis is perpendicular to the main tape plane. $I$-$V$ characteristics were measured by the 4-points method, using a Quantum Design Physical Property Measurement System (PPMS) as well as a home-made pulsed currents system in order to inject currents up to 50 A.

AC magnetic measurements were performed in the PPMS, using the so-called ACMS option. In any susceptometer, the presence of magnetic or conducting pieces in the vicinity of the sample may cause a distortion of the flux lines and significant phase errors [8]. This leads sometimes to the appearance of spurious peaks, as reported by a Quantum Design application note [9]. Therefore it is of interest to confront the results to those obtained with another system, especially when small signal variations are involved.

AC susceptibility measurements were thus also performed in a home-made susceptometer [10] enabling to apply parallel AC and DC magnetic fields which can be rotated around the sample. In order to directly investigate anisotropy effects in one measurement sequence, two orthogonal pick-up coils were wound around the tape and connected to an EGG 5210 lock-in amplifier. Each pick-up coil is made of fine (50 $\mu$m diameter) insulated copper wire tightly wrapped around the sample, in such a way that the magnetic flux through each measuring coil $\Phi$ is very close to the magnetic flux throughout the sample $\Phi_S$. When the magnetic field is applied in the $ab$-plane at 45° with respect to the filament direction, this method allows simultaneous measurements of the AC susceptibility components in two orthogonal directions.

The agreement of the data obtained by both experimental systems allows to attribute the results presented below to the intrinsic behavior of the tape.

Manuscript received August 5, 2002. This work was supported in part by the Région Wallonne (RW) through the VESUVE Contract.

J.-F. Fagnard and P. Vanderbemden are with the SUPRAS, Department of Electrical Engineering and Computer Science (Montefiore Institute, B28), University of Liege, B-4000 Liege, Belgium (e-mail: fagnard@montefiore.ulg.ac.be; Philippe.Vanderbemden@.ulg.ac.be).

R. Cloots is with the SUPRAS, Institute of Chemistry, University of Liege, B-4000 Liege, Belgium (e-mail: rcloots@ulg.ac.be).

M. Ausloos is with the SUPRAS, Institute of Physics, B5, University of Liege, B-4000 Liege, Belgium (e-mail: marcel.ausloos@ulg.ac.be).

Digital Object Identifier 10.1109/TASC.2003.812079

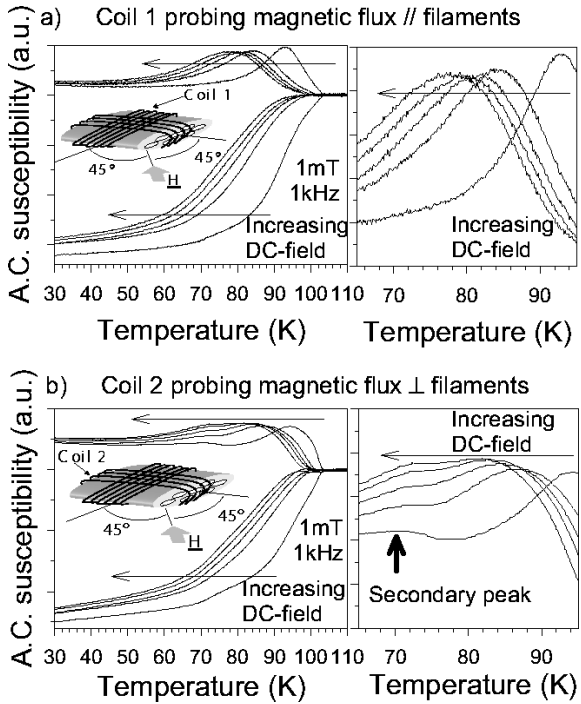

Fig. 1. AC susceptibility of Bi-2223/Ag tape as a function of the temperature with DC magnetic fields of 0, 25, 50, 75, and 100 mT parallel to the $ab$-plane at 45° with respect to the filament direction. The magnetic field is probed (a) parallel to the filament direction and (b) perpendicular to the filament direction.

## III. RESULTS AND DISCUSSION

### A. Transport Critical Current

First, the tape was electrically characterized at 77 K. The self-field transport critical current, extracted using a 1 $\mu$V/cm threshold, is 40 A, corresponding to a critical current density of about $1.6 \; 10^4$ A/cm$^2$. These results are typical characteristics of a good quality multifilamentary tape.

### B. AC Susceptibility Anisotropy

Fig. 1 shows the real and imaginary parts of the AC susceptibility measured using the home-made system described above. Several DC fields are superimposed to the 1 mT AC field ($H_{AC} \ll H_{DC}$). The fields are parallel to the $ab$-planes and directed at a 45° angle with respect to the filament direction. The two orthogonal pick-up coils scanning the AC magnetic response are sketched in the inset of Fig. 1. The first one (coil 1) has the turns perpendicular to the filaments and the second one (coil 2) has its turns parallel to the filaments. The right hand side of Fig. 1 shows also a blow-up of the out-of-phase susceptibility $\chi''$ around the peaks.

First we concentrate on the upper part of the Fig. 1 (flux component parallel to the filaments). The curves display the typical structure of a superconducting transition [11]. In the temperature range 60–100 K, the magnetic signal due to eddy currents in the silver is very small compared to that caused by critical currents flowing along the $ab$-planes ($J_{cJ}^{ab}$) and the $c$-axis ($J_{cJ}^{c}$) of the material. Since $H$ is parallel to the long axis of the filaments, the geometry is such that an anisotropic Bean model [12] can be used to analyze the data. In the present case, by assuming a priori $J_{cJ}^{c} \ll J_{cJ}^{ab}$, the full penetration field $H_p$ at a given temperature only depends on geometric factors and on the critical current parallel to the $c$-axis $J_{cJ}^{c}$. Strictly speaking, the position of the $\chi''$ peak depends on the exact sample geometry. By assuming each filament as an infinite slab, we can assume that, at the $\chi''$ peak temperature, $H_p$ is roughly equal to the applied AC field component parallel to the filaments multiplied by 3/4 [13]. This allows to obtain an order of magnitude for $J_{cJ}^{c}$, which is here estimated to be $2.9 \; 10^2$ A/cm$^2$ at 77 K and 70 mT. For comparison, at the same temperature and field, the transport $J_{cJ}^{ab}$ is $1.2 \; 10^4$ A/cm$^2$. This yields an anisotropy ratio $r = J_{cJ}^{ab}/J_{cJ}^{c}$ around 40. This value can be compared to those found in monocore tapes. By using DC magnetization, Cimberle et al. found 50 at $T = 50$ K [3], whereas Hensel et al. [14] found 10 at 77 K using a transport method. Note that all those values are much smaller than the intrinsic anisotropy ratio ($\sim 1000$) of the Bi-2223 phase [14].

Next we turn to the comparison of the results obtained for each of the sensing coils (upper and lower parts of Fig. 1). As can be seen, the magnetic transition component perpendicular to the filaments (bottom of Fig. 1) exhibits a particular structure: a *secondary* peak (marked with an arrow in Fig. 1) can be identified near 72 K and can be clearly distinguished from the main peak arising between 75 and 95 K. Measurements on several samples extracted from the same tape showed that this peak is always very small when the field is parallel to the filaments (and sometimes even not visible, as is the case in Fig. 1). Another striking feature observed in Fig. 1 is that the *secondary* peak position temperature does not change with the amplitude of the superimposed DC magnetic field.

It should be noticed that this peak never appears when $H || c$ (i.e., shielding currents flowing parallel to the $ab$-planes). It is, therefore, clearly linked to the component of the shielding currents flowing parallel to the $c$-axis. The fact that the secondary peak amplitude slightly varies from one specimen to another might suggest that it is caused by defects introduced during sample cutting. However the constant temperature range 71–73 K whatever the sample does rather suggest that the peak is the signature of some Bi-2212 intergrowth phase.

A double transition in the AC susceptibility of Bi-2223/Ag tapes has also been reported by Umezawa et al. [1], with the following characteristics differing from our results (i) their second transition occurs for $H || c$ and (ii) the temperature associated with their low-$T$ transition is 80 K. For monocore tapes, it has been shown that Bi-2212 intergrowths are located at twist boundaries between the Bi-2223 grains having common $c$-axis and forming a so-called grain colony [1], [14]. Since the secondary peak visible in Fig. 1 is linked to the $c$-axis current, i.e., the current crossing the $ab$-planes twist boundaries, it has clearly its origin in the presence of some Bi-2212-like intergrowth.

### C. AC Susceptibility Frequency Dependence

Now it is of interest to examine the conditions for which the low-$T$ peak is much more visible for $H$ perpendicular to the filaments than for $H$ parallel to the filaments although, in both cases, the field is parallel to the $ab$-planes and the shielding currents are expected to flow through the Bi-2212 intergrowths.

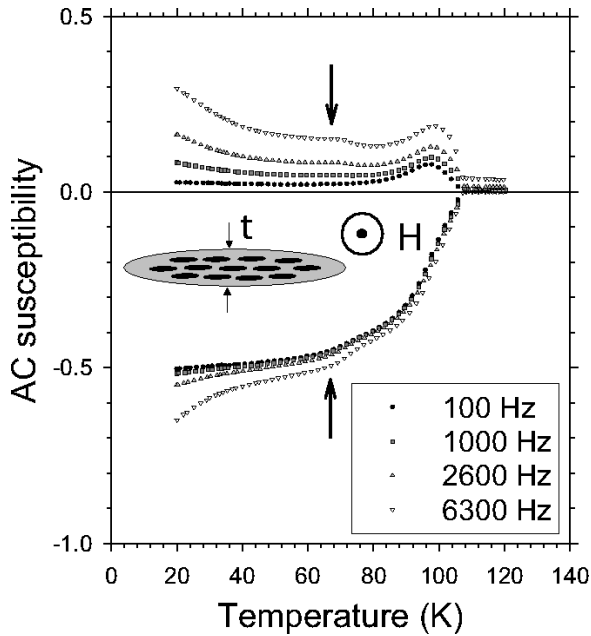

Fig. 2. AC susceptibility versus temperature, $H//$filaments ($\mu_0 H_{AC}$ = 1 mT). The arrows indicate the 72 K $\chi''$ peak and the 72 K $\chi'$ "foot" due to the Bi-2212 intergrowth.

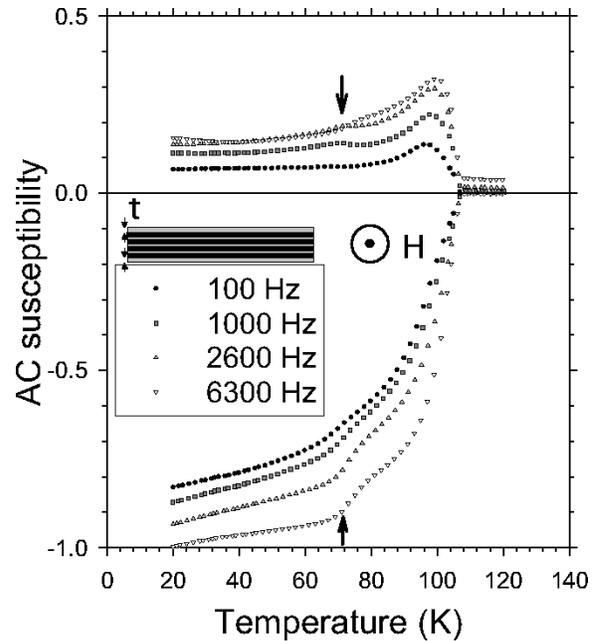

Fig. 3. AC susceptibility versus temperature, $H \perp$ filaments ($\mu_0 H_{AC}$ = 1 mT). The arrows indicate the 72 K $\chi''$ peak and the 72 K $\chi'$ "foot" due to the Bi-2212 intergrowth.

Measurements were performed at several frequencies and AC magnetic field amplitudes. Results are presented in Fig. 2 ($H$||filaments) and Fig. 3 ($H \perp$ filaments) for 1 mT and frequencies ranging from 100 to 6300 Hz.

First of all, we examine the temperature range $T < 50$ K. In this regime, the resistivity of the Ag matrix is small and at high frequencies, the signal is expected to be dominated by eddy currents in the silver. This is clearly the case when the magnetic field is parallel to the filaments (Fig. 2) because the curvature of the $\chi' - \chi''$ vs. $T$ curves exactly mimics what was measured on a silver sheet having approximately the same thickness (0.18 mm) as the superconducting tape investigated here. For a magnetic field directed perpendicular to the filament direction (Fig. 3), the situation is rather different: the in-phase susceptibility ($\chi'$) linearly decreases with decreasing temperature and the out-of-phase component ($\chi''$) remains almost temperature-independent, whatever the frequency.

It is also worth comparing the low-temperature values of the AC susceptibility shown in Figs. 2 and 3. For $H$||filaments (Fig. 2) and low frequencies, the AC susceptibility nearly saturates at $-0.5$, which is very close to the true fraction of superconductor in the tape cross-section. On the contrary, for $H \perp$ filaments (Fig. 3), the demagnetization factor of the filaments (here of the order of 0.3–0.4) should be taken into account, giving thus a susceptibility which is, in absolute value, larger than the true superconducting fraction.

Next, we examine the temperature range 50 – 110 K. As can be seen in Figs. 2 and 3, the temperatures at which the main ($T \sim 100$ K) and the secondary ($T \sim 72$ K) superconducting transitions occur are nearly frequency independent. For both magnetic field geometries, increasing the AC field frequency enhances the $\chi'$ drop around 72 K. However, when $H \perp$ filaments, this enhancement is much more pronounced while the associate peak seems to disappear (Fig. 3).

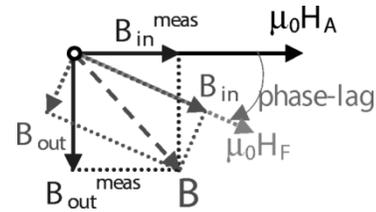

Fig. 4. Vectorial diagram showing the phase difference between the applied magnetic field $H_A$ and the local magnetic field seen by the superconducting filament, $H_F$. The induction $B$ has two components, respectively in-phase ($B_{in}$) and out-of-phase ($B_{out}$) with respect to $H_F$. For a given value of $B$, the measured in-phase ($B_{in}^{meas}$) and out-of-phase ($B_{out}^{meas}$) components are referred to $H_A$.

This behavior finds a reasonable explanation by considering that, with increasing frequency, the eddy currents in the silver induce an increasing significant phase difference between the *applied* magnetic field, $H_A$, and the *internal* magnetic field seen by the filaments, $H_F$. Of course flux flow mechanisms cannot be excluded but the frequency dependence of such mechanisms are believed to have much less impact on both the $\chi''$ peak amplitude and the $\chi'$ value that what is observed in our measurements. Moreover, the significant value of $\chi''$ above $T_c$ at the highest measurement frequency shows the influence of eddy currents on the measurements. A frequency-dependent phase difference between $H_A$ and $H_F$ has important consequences for the measured signal, and in particular for the conditions at which the secondary $\chi''$ peak can be observed. This is schematically shown in the vectorial diagram of Fig. 4. Due to eddy currents, the field seen by the filaments, $H_F$, is phase-lagged with respect to the applied field $H_A$ and due to the superconducting losses, the induction $B$ can be decomposed into two components, respectively in-phase ($B_{in}$) and out-of-phase ($B_{out}$) with respect to $H_F$. Such a situation affects the measured in-phase ($B_{in}^{meas}$) and out-of-phase ($B_{out}^{meas}$) components, referred to $H_A$. More precisely, as frequency increases, the secondary peak connected to the losses in the superconductor should progressively appear in the measured in-phase susceptibility component ($\chi'$) and disappear from the measured out-of-phase component ($\chi''$). This is exactly what can be seen in Fig. 3: the abrupt shielding enhancement at 72 K corresponds to a 'transfer' of the peak from $\chi''$ to $\chi'$. Moreover, as frequency increases, the increasing phase difference between $H_A$ and $H_F$ implies that the *measured* induction component in-phase with the applied field becomes progressively smaller than its true effective value (cf. Fig. 4). Therefore the *measured* in-phase susceptibility should increase (in absolute value) with frequency even in the case of a frequency independent superconducting shielding. This is clearly the behavior observed in Fig. 3.

From the above results, it can be concluded that the field configuration where $H$ ($\|ab$) $\perp$ filaments induces a situation where the phase difference between $H_A$ and $H_F$ significantly affect the measurements. This is not the case when $H$ ($\|ab$) is directed parallel to the filaments.

## IV. Conclusion

We have experimentally studied the AC susceptibility of multifilamentary Bi-2223/Ag tapes for $H\|ab$. The results put into evidence the two following anisotropic features: (i) a secondary peak appearing near 72 K and (ii) the influence of the phase difference between the *applied* and the *local* field seen by the filaments on the susceptibility measurements. These characteristics are strongly enhanced for $H \perp$ filaments.

## Acknowledgment

The authors would like to acknowledge Prof. H. W. Vanderschueren for allowing the use of MIEL laboratory facilities. Thanks also go to RW, FNRS, and ULg for cryofluid and equipment grants. One of the authors, P. Vanderbemden, is the recipient of a FNRS post doctoral fellowship.